\documentclass[prl,twocolumn,showpacs,amsmath,amssymb]{revtex4}
\usepackage{graphicx}
\usepackage{indentfirst}
\usepackage{psfrag}
\usepackage{epsfig}
\usepackage{amsmath}
\usepackage{amssymb}
\usepackage{bm}
\usepackage{mathrsfs}
\newcommand{\be}{\begin{eqnarray}}
\newcommand{\ee}{\end{eqnarray}}

\newcommand{\hel}{\mathscr{ H}}
\newcommand{\leh}{\mathscr{ L}}
\newcommand{\flow} {\mathscr{ F}}
\newcommand{\perd}{\mathscr{ P}}
\newcommand{\qui}{\mathscr{ K}}
\newcommand{\s}{\mathscr{ S}}

\begin{document}

\title{Optical theorem for the conservation of electromagnetic helicity: Significance for molecular energy transfer and enantiomeric discrimination by circular dichroism} 
\author{Manuel Nieto-Vesperinas }
\affiliation{Instituto de Ciencia de Materiales de Madrid, Consejo Superior de
Investigaciones Cient\'{i}ficas\\
 Campus de Cantoblanco, Madrid 28049, Spain.\\ www.icmm.csic.es/mnv; 
mnieto@icmm.csic.es }

%\author{}
%\authors{

%\email{mnieto@icmm.csic.es}

%\affiliation{
\begin{abstract}
We put forward the physical meaning of the conservation equation for the   helicity  on scattering of an electromagnetic field with a generally magnetodielectric  bi-isotropic dipolar object. This is the optical theorem for the helicity that, as we find, plays  a role for this quantity analogous to that of the optical theorem for energy.  We discuss its consequences for helicity transfer between molecules and for new detection procedures of circular dichroism based on  ellipsometric measurements.
\end{abstract}
\maketitle

 \section{Introduction}

Several effects derived from the twisting of the polarization and wavefronts of electromagnetic fields, specifically the spin and orbital angular momenta, are  a subject of increasing study in  recent years  \cite{allenlibro,allenlibro1,babiker,yao,cameron2,molina1,molina2}. This is accompanied by a steady improvement in particle manipulation techniques and theories  \cite{garces,grier,dunlop,chaumet,brasse,chen,MNV2015}, and by  the use of spatially structured waves \cite{barron2} with enhanced helicity \cite{tang1} to increase the signal in circular dichroism \cite{schellmann,barron1}  for enantiomeric discrimination \cite{tang2, tang3,choi}. In addition, recent studies \cite{muka} in fluorescence resonance energy transfer (FRET) \cite{fret1,fret2}, (see also \cite{craig,salam}),  show an electromagnetic force between excited molecules, different from the Van der Waals force when they  are in their ground-state.

 A consequence of this research was the derivation of a  conservation law for the helicity of electromagnetic fields \cite{lipkin,cameron1} that appears as fundamental as that for the energy.

In this paper we discuss the physical significance of this helicity conservation   law.  Dealing  with quasimonochromatic electromagnetic fields, we establish the optical theorem which constitutes the main consequence of this law concerning  optical, or electromagnetic, scattering. In this way, we show that this new theorem  provides an expression for the helicity excitation rate of a particle, (dipolar in the wide sense, i.e. such that its scattering may be fully described by its first electric and magnetic partial waves), by extinction of  the helicity of the irradiating field. In particular for magnetodielectric bi-isotropic objects,  this leads to a relationship  between polarizabilities, complementary and compatible with that of the optical theorem for energies. For circularly polarized light this also  establishes a necessary and  sufficient condition between their  duality and  scattering characteristics. 

 In this respect, we do not address here quadrupoles or other multipolar excitations. Although extensions of dipolar models have been carried out in studies of  the energy conveyed by those higher order terms, showing the observable signal due to the electric dipole-quadrupole polarizability for chiral configurations \cite{Yang}, (see also \cite{choi} remarking the similarity in magnitude of the electric quadrupole and magnetic dipole moments according to  quantum electrodynamical calculations in \cite{craig, salam}),
as regards the purpose of our study which deals  with a  different quantity: the helicity, we show  that the (broad sense) dipolar formulation already leads to new physical phenomena that should be observed in future novel experiments, even though of course this theory  is amenable of further generalizations to account for effects due to higher order excitations.

More importantly, this novel equation opens a new landscape for:

1.The emission and absorption of helicity in complex environments, also  in particular  at the nanoscale, e.g. in FRET between molecules, or other nanoscructures, where rather than anlysing the transference of energy, one establishes and addresses the behavior of  the helicity  lifetimes, taking the  bi-isotropy, and chirality in particular, into account.

2. Enantiomeric discrimination, where chiral molecules, or other nanoparticles, are studied by circular dichroism. This is done  by means of a new dissymmetry  factor introduced in this  work stemming from this novel  optical theorem. This factor has higher sensitivity than the standard one \cite{barron1}  based  on the extinction of incident energy  and its transfer to the object by  measuring its intensity excitation, since it involves a new experimental procedure  which detects the total scattered helicity and its flow by means of  an ellipsometry set-up \cite{marston1}.

\section{The helicity}

We consider  fields, currents and potentials  with a time-harmonic dependence, so that the electric and magnetic vectors are   ${\bf \cal E}({\bf r},t)$ and  ${\bf \cal B}({\bf r},t)$:  ${\bf \cal E}({\bf r},t)=\Re [{\bf E}({\bf r}) \exp(-i\omega t)]$ and  ${\bf \cal B}({\bf r},t)=\Re [{\bf B}({\bf r}) \exp(-i \omega t)]$.  $\Re$ denotes real part.

We introduce the helicity density $\mathscr{ H}$ and the density of flow of helicity  $\mathscr{ F}$  of this  field in a non-absorbing dielectric medium of  refractive index $n=\sqrt{\epsilon \mu}$, ($\epsilon$ and $\mu$ represent the dielectric permittivity and the magnetic permeability),   as:
\begin{equation}
\hel=\frac{1}{2}(\frac{1}{\mu}{\bf \cal A}\cdot\bf{\cal B}-\epsilon {\bf \cal C} \cdot \bf \cal E), \label{h}
\end{equation}
\begin{equation}
\flow=\frac{c}{2 \mu}({\bf \cal E}\times\bf{\cal A} +  {\bf \cal B} \times \bf \cal C). \label{f}
\end{equation}
 ${\bf \cal A}$ and ${\bf \cal C}$ are vector potentials  such that: ${\bf \cal B}=\nabla \times {\bf \cal A}$ and  ${\bf \cal E}=-\nabla \times {\bf \cal C}$
\cite{cameron1}, so that  working in a Coulomb gauge: $\nabla\cdot {\bf \cal A}=\nabla\cdot {\bf \cal C}= 0$, and  one has from Maxwell's equations: 
\be
\dot{\bf \cal A}= -c {\bf\cal  E},
\dot{\bf \cal C}= - \frac{c}{\epsilon\mu} \nabla\times {\bf \cal A } + \frac{4 \pi}{\epsilon}  {\bf \cal K };
 {\bf \cal J}= \nabla \times {\bf \cal K }. \label{pots}
\ee

The upper dot stands for $\partial_t$, $c$ is the light speed in vacuum, and $\bf \cal J$ denotes the electric current density which is transversal since the existence of ${\bf \cal A}$ and the law $\nabla\cdot \epsilon {\bf \cal E}= 4 \pi \rho$ imply that the electric charge density $\rho$ is zero . From the above equations one obtains the conservation law \cite{cameron1}
\begin{equation}
\dot{\hel}+ \nabla \cdot \flow = - \perd . \label{contH}
\end{equation}
Where dissipation in the  interaction of the fields with matter is represented by $\perd= 2\pi ({\bf  \cal E}\cdot {\bf  \cal K}- {\bf \cal J} \cdot {\bf \cal  C})$.

Since the fields and potentials  are time-harmonic, we convert the quantities holding Eqs. (\ref{pots})  and  (\ref{contH}) into:
\be
{\bf  A}= - \frac{i}{k} {\bf E}, \,\,\,\,
{\bf  C}= - \frac{i}{\epsilon} [\frac{{\bf B}}{k \mu} - \frac{4 \pi}{\omega} {\bf K}],  \label{potsr}
\ee
and
\begin{equation}
\hel= <\hel>=\frac{1}{2k}\sqrt{\frac{\epsilon}{\mu}} \Im ({\bf E}\cdot {\bf B}^*), \label{hh}
\end{equation}
\begin{equation}
\flow= <\flow> =  \frac{c}{4n k } \Im ( \epsilon {\bf E}^* \times {\bf  E}  + \frac{1}{\mu}{\bf  B}^*\times \bf B)  . \label{ff}
\end{equation}
Where $<\cdot>$ denotes time-average, $\Im$  means imaginary part and $k=n \omega/c$, ${\bf \cal A}= \Re[ {\bf A}({\bf r}) \exp (-i \omega t)]$, 
${\bf \cal C}= \Re[{\bf C}({\bf r})  \exp (-i \omega t)]$,   ${\bf \cal J}= \Re[{\bf J}({\bf r})  \exp(-i \omega t)]$,  ${\bf \cal K}=\Re[ {\bf K}({\bf r})  \exp(-i \omega t)]$.  Now $\flow$ coincides with the spin angular momentum density. Eq.(\ref{contH}) is then fulfilled by these time-averaged quantities  with $\perd$  replaced by:
\begin{eqnarray}
< \perd >=\pi [\frac{2}{ck}\sqrt{\frac{\mu}{\epsilon}} \nabla\cdot \Im ({\bf K}\times {\bf B}^*)  \nonumber \\ -\frac{1}{kn} \Im ( {\bf J} \cdot {\bf  B}^*)  +\frac{4\pi}{ck}\sqrt{\frac{\mu}{\epsilon}}\Im  ({\bf J}\cdot {\bf K}^*)  
 +\Re  ({\bf E}\cdot {\bf K}^*) ]  . \label{pp}
\end{eqnarray}
For these monochromatic fields, Maxwell's  equations, and the above relations, show   that (\ref{hh}) and (\ref{ff}) are proportional to Lipkin's zilches \cite{lipkin,cameron1}, used in recent works as chirality $\qui$  and flow of chirality $\s$ \cite{tang1,tang2}:
\begin{equation}
\qui= <\qui>=k^2 \hel= k^2 <\hel>\label{kk}
\end{equation}
\begin{equation}
\s= <\s> = k^2 \flow= k^2 <\flow>\label{ss}
\end{equation}
The dissipative terms are however different. 
We follow the criterion  of \cite{cameron1} according to which $\flow$ is the quantity with dimensions of angular momentum, so that we deal with  the helicity and its flow; although  (\ref{kk}) and (\ref{ss}) show that both pairs yield equivalent  mesurements   for monochromatic fields.
\section{The optical theorem for the helicity}
Let a  monochromatic, elliptically polarized, plane wave be incident on a scattering body, e.g. a polarizable particle, (cf. Fig.1). The field at any point of the exterior medium may be represented as the sum of the incident  and the scattered vectors  as:  ${\bf E}({\bf r})={\bf E}_i({\bf r})+{\bf E}_s({\bf r})$,   ${\bf B}({\bf r})={\bf B}_i({\bf r})+ {\bf B}_s({\bf r})$. 

\begin{figure}[htbp]
\centerline{\includegraphics[width=1.0\columnwidth]{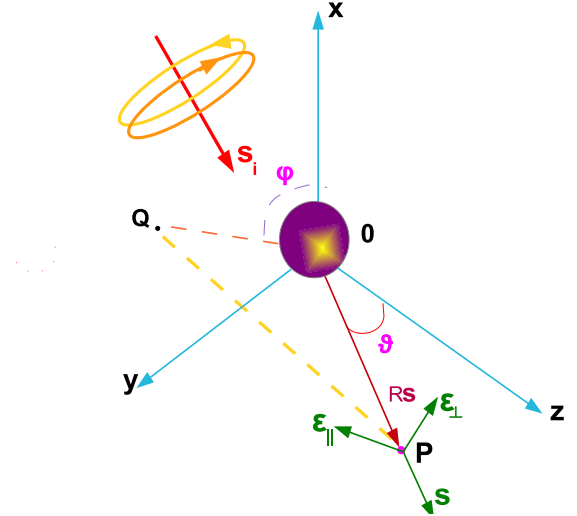}}
\caption{(Color online). An elliptically polarized plane wave incides on a  polarizable particle. The fields are evaluated at the point {\bf P}: ${\bf r}=R \bf s$ , of  coordinates   $(R, \theta, \phi)$, of  a sphere of integration   of radius $R$, centered at some point  ${\bf r}_ {0}$  of the particle.  ${\bf r}_ {0}$ acts as the framework center {\bf 0}. The point {\bf Q} is the projection of {\bf P} on the plane $ OXY$; the scattering plane being $OPQ$.  We  show the three orthonormal vectors:  ${\bf s}$,  ${\bm \epsilon}_{\parallel}$ (in the plane $OPQ$ and in the sense of rotation of $\theta$), and ${\bm \epsilon}_{\perp}$ (normal to $OPQ$).}
\end{figure}

The incident fields being  ${\bf E}_i={\bf e}_{i} e^{ik({\bf s}_{i}\cdot {\bf r})}$,  ${\bf B}_i={\bf b}_{i} e^{ik({\bf s}_{i}\cdot {\bf r})}$; whereas in  the far zone the scattered fields are:  ${\bf E}_s={\bf e}({\bf s}) \exp(ikr)/r$, ${\bf B}_s={\bf b}({\bf s}) \exp(ikr)/r$ . Also  ${\bf b}_{i}=n {\bf s}_{i} \times  {\bf e}_{i}$, ${\bf e}_{i} \cdot {\bf s}_{i}= {\bf b}_{i} \cdot {\bf s}_{i}=0$;  ${\bf b}=n {\bf s} \times  {\bf e}$, ${\bf e} \cdot {\bf s}= {\bf b} \cdot {\bf s}=0$.

 The flow (or time-averaged flow) density of helicity is:  $\flow= <\flow> =  \flow^i + \flow^s +
\flow' $  . Where
\begin{eqnarray}
\flow^i=<\flow^i>=\frac{c}{4n k } \Im ( \epsilon {\bf E}_{i}^{ *} \times {\bf  E}_{i }  + \frac{1}{\mu}{\bf  B}_{i}^{ *}\times {\bf B}_{i })  .   (10a)  \nonumber\\
\flow^s=<\flow^s>=\frac{c}{4n k } \Im ( \epsilon {\bf E}_{s}^{ *} \times {\bf  E}_{s }  + \frac{1}{\mu}{\bf  B}_{s}^{ *}\times {\bf B}_{s}).  \nonumber (10b) \\
\flow'=<\flow'>=\frac{c}{4n k } \Im ( \epsilon {\bf E}_{i}^{ *} \times {\bf  E}_{s }  + \frac{1}{\mu}{\bf  B}_{i}^{ *}\times {\bf B}_{s }).  \nonumber (10c)
\end{eqnarray}

From Eq.(\ref{contH}) the rate ${\cal W}_{\hel}^{a}$ at which the helicity is dissipated on interaction with the body is given by the $\Sigma$-integral that gives the outward flow of helicity: $\int_{\Sigma}  d\Omega R^2 \flow  \cdot {\bf s}$   through the surface of a large sphere $\Sigma$ of radius $R$ with center at some point  ${\bf r}_0$ of the object. $d\Omega$ is the element of solid angle  and ${\bf s}$ denotes the outward normal.  I.e., according to Eq.(\ref{contH}):
\begin{equation}
- {\cal W}_{\hel}^{a}= {\cal W}_{\hel}^{i}+{\cal W}_{\hel}^{s}+{\cal W}_{\hel}'. \label{who}
\end{equation}
Where ${\cal W}_{\hel}^{i}$, ${\cal W}_{\hel}^{s}$ and ${\cal W}_{\hel}'$ are respectively the  $\Sigma$-integrals of the projections on $\bf s$ of  $\flow^i$, $\flow^s$ and $\flow' $.  On the other hand,  ${\cal W}_{\hel}^{a}=\int_{\Sigma} d R d\Omega R^2 <\perd >$.

From these equations we have that  $ {\cal W}_{\hel}^{i}=0$, so that  (\ref{who}) becomes
\be
 {\cal W}_{\hel}^{a}+ {\cal W}_{\hel}^{s}=-{\cal W}_{\hel}'. \label{wh}
\ee
Whereas the integrals  of $\flow^s \cdot {\bf s}$ and $\flow' \cdot {\bf s}$ across $\Sigma$  are 
\be
{\cal W}_{\hel}^{s}=\int_{\Sigma}  d\Omega R^2   \flow^{s} \cdot {\bf s} =\nonumber \\
\frac{c}{4n k } \Im \int_{\Sigma}  d\Omega {\bf s} \cdot [ \epsilon {\bf e}^{*}({\bf s}) \times  {\bf e}({\bf s}) + \frac{1}{\mu}{\bf b}^{*}({\bf s}) \times  {\bf b}({\bf s})] , \label{fss}
\ee
and
\be
{\cal W}_{\hel}'=\int_{\Sigma}  d\Omega R^2   \flow'    \cdot {\bf s}=\nonumber \\
\frac{2 \pi c }{n  k^2 }\sqrt{\frac{\epsilon}{\mu}} \Re[{\bf B}_{i}^{*}(\bf r_0)   \cdot  {\bf e}({\bf s}_{i})] = \nonumber \\
-\frac{2 \pi c }{n  k^2 }\sqrt{\frac{\epsilon}{\mu}} \Re[{\bf E}_{i}^{*}(\bf r_0)  \cdot  {\bf b}({\bf s}_{i})] . \label{fis}
\ee
In deriving (\ref{fis}) we have used Jones' lemma based on the principle of the stationary phase \cite{jones, born}:

\be
\frac{1}{R}\int  d\Omega R^2  F({\bf s}) e^{-ik ({\bf s}_{i} \cdot {\bf s})R} \sim \frac{2\pi i}{k}[F({\bf s}_{i})e^{-ik R} \nonumber \\
 - F(-{\bf s}_{i})e^{ik R} ] . \label{psp}
\ee
Eqs.(\ref{fss}) and  (\ref{fis})  together with (\ref{wh})  constitute the {\it optical theorem} that represents the {\it conservation of helicity} on scattering by an arbitrary body. They state that the rate at which the helicity is dissipated from the incident wave, in the form of losses in the obstacle,  ${\cal W}_{\hel}^{a}$, and  of  helicity of the scattered field integrated in all directions, ${\cal W}_{\hel}^{s}$, is proportional to a certain helicity component of the scattered field, $-{\cal W}_{\hel}'$,  on interference with the incident wave in the forward direction ${\bf s}_i$. This suggests to  introduce a {\it helicity extinction cross-section} of the body ${\cal Q}_{\hel}$  dividing the terms of (\ref{wh}) by  the rate $|\flow^{i}|=\frac{c}{2nk}\sqrt{\frac{\epsilon}{\mu}}\Im [{\bf E}_{i} ({\bf r}_0) \cdot {\bf B}_{i}^{*}({\bf r}_{0}) ]=\frac{c}{2nk}\sqrt{\frac{\epsilon}{\mu}}{\Im [{\bf e}_{i} \cdot {\bf b}_{i}^{*}]} $ at which the helicity is incident on a unit cross-sectional area of the object, so that (\ref{wh})  and (\ref{fis})   give:
\be
{\cal Q}_{\hel}=\frac{{\cal W}_{\hel}^{a}+{\cal W}_{\hel}^{s}}{|\flow^{i}|}=\nonumber\\
-\frac{4 \pi }{k}\frac{\Re[{\bf B}_{i}^{*}(\bf r_0)   \cdot  {\bf e}({\bf s}_{i})]} {\Im [{\bf e}_{i } \cdot {\bf b}_{i}^{*} ]}= \frac{4\pi }{k}\frac{\Re[{\bf E}_{i}^{*}(\bf r_0)   \cdot  {\bf b}({\bf s}_{i})]}{\Im [{\bf e}_{i}  \cdot {\bf b}_{i}^{*} ]}. \label{ot}
\ee
Where $\Im [{\bf E}_{i} \cdot {\bf B}_{i}^{*}]=\Im [{\bf e}_{i} \cdot {\bf b}_{i}^{*}]=2k \sqrt{ \mu/ \epsilon} {\hel^{i}}$ .   $ {\hel^{i}}$ being  the helicity of the incident wave.    For this field  $\hel^{i}=(n/c)\flow^{i} \cdot {\bf s}_{i}$.

Eq.(\ref{ot})  is the {\it optical theorem} for the {\it helicity cross-section}. Likewise,  the {\it  absorption} and {\it scattering helicity cross-sections} ${\cal Q}_{\hel}^{a}$ and ${\cal Q}_{\hel}^{s}$ are  introduced as:
\be
{\cal Q}_{\hel}^{a}=\frac{{\cal W}_{\hel}^{a}}{|\flow^{i}|} , \,\,\,\,\,
{\cal Q}_{\hel}^{s}=\frac{{\cal W}_{\hel}^{s}}{|\flow^{i}|}. \label{qs}
\ee
And of course ${\cal Q}_{\hel}={\cal Q}_{\hel}^{a}+{\cal Q}_{\hel}^{s}$.

These laws embody a close analogy with those of the optical theorem for the energy \cite{born}, and suggest the determination of these magnitudes in scattering experiments.

\section{Magnetodielectric bi-isotropic dipolar particle}
Let us consider a  magnetodielectric bi-isotropic particle \cite{kong}, dipolar in the wide sense, i.e.  if for example we consider it a sphere, its  magnetodielectric response, is characterized by its electric,  magnetic, and magnetoelectric polarizabilities $\alpha_{e}$,  $\alpha_{m}$,  $\alpha_{em}$,  $\alpha_{me}$, given by the first order Mie coefficients as: $
\alpha_{e}=i\frac{3}{2k^{3}}a_{1}$,  $\alpha_{m}=i\frac{3}{2k^{3}}b_{1}$,  $\alpha_{em}=i\frac{3}{2k^{3}}c_{1}$,  $\alpha_{me}=i\frac{3}{2k^{3}}d_{1}=-\alpha_{em}$.  $a_{1}$, $b_{1}$ and  $c_{1}=-d_{1}$ standing for the electric, magnetic, and magnetoelectric first Mie coefficients, respectively \cite{MNV2010,Nieto2011,chanlat}. Notice that such sphere is chiral since   $\alpha_{em}=-\alpha_{me}$. 

The electric and magnetic dipole moments, $ {\bf p}$ and  ${\bf m}$, induced on the particle by the incident field are:
\be
{\bf p}=\alpha_{e} {\bf E}_i+\alpha_{em}{\bf  B}_i, \,\,\,\,\,
{\bf m}=\alpha_{me}{\bf E}_i+\alpha_{m}{\bf B}_i. \label{consti}
\ee
And the fields scattered by this particle  in the far-zone read:
\begin{equation}
{\bf e}({\bf s})=k^2 \frac{e^{ikr}}{r} [\epsilon^{-1}({\bf s} \times {\bf
p})\times {\bf s}- \sqrt{\frac{\mu}{\epsilon}}({\bf s}\times
{\bf m})], \label{dipe}
\end{equation}
\begin{equation}
{\bf b}({\bf s})=k^2 \frac{e^{ikr}}{r}[\mu({\bf s} \times {\bf m})\times
{\bf s}+\sqrt{\frac{\mu}{\epsilon}} ({\bf s}\times {\bf p})]. \label{dipm}
\end{equation}
Introducing (\ref{dipe}) and (\ref{dipm}) into (\ref{fss}) and (\ref{fis}), evaluating the angular integrals, and substituting the results in (\ref{wh}) we obtain:
\be
 {\cal W}_{\hel}^{a}+\frac{8\pi c k^3}{3 \epsilon} \Im[{\bf p} \cdot {\bf m}^{*}]= \nonumber \\
\frac{2\pi c}{n} \{ \Re[ ({\bf p}  \times {\bf E}_{i}^{*})    \cdot{\bf s}_{i}] + \Re[({\bf m}\times  {\bf B}_{i}^{*}) \cdot{\bf s}_{i} ] \}. \label{todip1}
\ee
Which normalizing to ${|\flow^{i}|}$, becomes {\it  the optical theorem for the helicity}  expressed as:
\be
 {\cal Q}_{\hel}^{a}+\frac{4\pi}{3}  \frac{k^3}{\hel^{i}} \sqrt{\frac{\mu}{\epsilon} } \Im[{\bf p} \cdot {\bf m}^{*}]= \nonumber \\
\frac{\pi }{\hel^{i}} \{ \Re[({\bf p} \times {\bf E}_{i}^{*}) \cdot{\bf s}_{i} ]+ \Re[({\bf m}  \times {\bf B}_{i}^{*})    \cdot{\bf s}_{i}] \}. \label{todip}
\ee

The second term of  the left side of  (\ref {todip}) is the "total scattered helicity cross section"  or  helicity  scattering-cross section defined above, [cf. (\ref{qs})],
\be
{\cal Q}_{\hel}^{s}=\frac{4\pi}{3}  \frac{k^3}{\hel^{i}} \sqrt{\frac{\mu}{\epsilon} } \Im[{\bf p} \cdot {\bf m}^{*}],   \label{qs1}
\ee
associated to the rate of helicity excitation;  as such it accounts for optical rotation effects like e.g. {\it circular dichroism}  \cite{schellmann,barron1}. 

On the other hand, the right side of  (\ref {todip1}) or (\ref {todip}) is proportional to the projection on ${\bf s}_i$ of the extinction optical torque  $\bm \Gamma$ felt by the particle \cite{MNV2015}:
\be
{\bm \Gamma}=
\frac{1 }{2}  \Re \{[ {\bf p}\times {\bf E}_{i}^{*} ] + [ {\bf m}\times{\bf B}_{i}^{*} ] \} , \label{torque}
\ee
 exerted by the spin of the incident wave.  Notice also that since $({\bf p} \times {\bf E}_{i}^{*}) \cdot{\bf s}_{i}= -(1/n) {\bf p}  \cdot {\bf B}_{i}^{*}$ and  $({\bf m} \times {\bf B}_{i}^{*}) \cdot{\bf s}_{i}= n {\bf m}  \cdot {\bf E}_{i}^{*}$, this term may also be expressed as $
\frac{2\pi c}{\mu} \Re \{ -\frac{1}{\epsilon} {\bf p}  \cdot {\bf B}_{i}^{*} +\mu  {\bf m}\cdot  {\bf E}_{i}^{*}  \}$, therefore the {\it conservation of helicity}  (\ref{todip1}) may be written as
\be
{\cal W}_{\hel}^{a}+\frac{8\pi c k^3}{3 \epsilon} \Im[{\bf p} \cdot {\bf m}^{*}]= 
\frac{2\pi c}{\mu} \Re \{ -\frac{1}{\epsilon} {\bf p}  \cdot {\bf B}_{i}^{*} +\mu  {\bf m}\cdot  {\bf E}_{i}^{*}  \}. \,\,\,\,\label{todip4}
\ee

The condition (\ref{todip4}) must be compatible with the optical theorem for energies \cite{born}
\be
 {\cal W}^{a}+\frac{c k^4}{3 n} [\epsilon^{-1} |{\bf p}|^2 + \mu |{\bf m}|^{2}]= \nonumber \\
\frac{\omega}{2} \Im[ {\bf p}  \cdot {\bf E}_{i}^{*}   +{\bf m}\cdot  {\bf B}_{i}^{*}] . \label{top}
\ee
${\cal W}^{a}$ being the rate of energy absorption, the second term of the left side constituting the total energy scattered by the dipolar object, and the right side representing the energy rate dissipated  from the illuminting field.  In this connection, notice the interesting formal analogy in the two conservation laws (\ref{todip4}) and (\ref{top}) where we observe a duality of  ${\bf E}$  and ${\bf B}$.  Also  the comparison between the second terms of their respective left sides is intriguing. We shall  discuss these points in the  next section. 

Notwithstanding let us remark that for circularly polarized light,  when we add  the two scattering  cross sections, namely, that of helicity: ${\cal Q}_{\hel}^{s}$ given by Eq. (\ref{qs1}), and that of  energy:  ${\cal Q}=\frac{c k^4}{3 n}[\epsilon^{-1} |{\bf p}|^2 + \mu |{\bf m}|^{2}]/|<{\bf S}_i>|$, (the denominator is the incident energy flow magnitude), then {\it  the product}: $[{\cal Q}_{\hel}^{s}+{\cal Q}]{\cal W}_i$, (where ${\cal W}_i$ is the  incident illuminating energy density), {\it represents the  rate  of excitation} of a chiral molecule or particle. An expression  usually derived from  quantum mechanics \cite{craig} and that here we have obtained on the basis of Maxwell's equations.

The conservation of helicity is generalized to an arbitrary illuminating wavefield, which we express as a decomposition of plane wave components \cite{mandel, nietolib}: 
\be
{\bf E}^{(i)}({\bf r})=\int_{\cal D} {\bf e}_{i}({\bf s}) e^{ik({\bf s}\cdot {\bf r})} d\Omega, \nonumber \\
{\bf B}^{(i)}({\bf r})=\int_{\cal D} {\bf b}_{i}({\bf s}) e^{ik({\bf s}\cdot {\bf r})} d\Omega.  \label{ang1}
\ee
The integration being done in the  contour $\cal D$ that contains both propagating and evanescent waves \cite{mandel,nietolib},   and to include them  both, ${\bf s}_i$ in  (\ref{todip1})  and (\ref{todip}) must be replaced by  ${\bf s}_{i}^{*}$, complex conjugated  of  ${\bf s}_{i}=(s_{i}^{x},
s_{i}^{y}, s_{i}^{z}) $, where  $s_{i}^{z}= \sqrt{ 1-(s_{i}^{x 2}+s_{i}^{y 2})}$  if  $ s_{i}^{x 2}+s_{i}^{y 2} \leq 1$, (propagating components); and   $s_{i}^{z}= i \sqrt{ (s_{i}^{x 2}+s_{i}^{y 2})-1}$ if   $s_{i}^{x 2}+s_{i}^{y 2} > 1$, (evanescent  components). Then by the same procedure as before and summing up for all plane wave components,  one sees that in Eqs.(\ref{todip1})-(\ref{todip4}) now ${\bf  E}_{i}$ and   ${\bf  B}_{i}$  must replaced by  ${\bf  E}^{(i)}$ and  ${\bf  B}^{(i)}$;  therefore instead of  (\ref{todip4}) we now obtain the  fundamental  {\it conservation relation for the helicity}:
\be
 {\cal W}_{\hel}^{a}+\frac{8\pi c k^3}{3 \epsilon} \Im[{\bf p} \cdot {\bf m}^{*}]= \nonumber \\
2\pi c \Re \{ -\frac{1}{n^2} {\bf p}  \cdot {\bf B}^{(i) *} + {\bf m}\cdot  {\bf E}^{(i) *}  \}. \label{todip3}
\ee
We remark that in this general case the incident fields in the right side of  (\ref{top}) should also be ${\bf E}^{(i)}$ and  $ {\bf B}^{(i)}$.

Equations (\ref{todip1}),  (\ref{todip}) or (\ref{todip4}), as well as Eq. (\ref{todip3}), express the extinction of helicity from the incident field by interaction with the dipolar particle. The right side is the helicity dissipated by the dipole from the illuminating wave, and plays for this magnitude a role analogous to that of  $\frac{\omega}{2} \Im[{\bf p} \cdot  {\bf E}^{(i) *} +  {\bf m}    \cdot {\bf B}^{(i)  *}]  $ for the dissipated energy. As such, the term $2\pi c \Re \{ -\frac{1}{n^2} {\bf p}  \cdot {\bf B}^{(i) *} + {\bf m}\cdot  {\bf E}^{(i) *}  \}$  has a potential for determining both dissipated and radiated, or scattered, helicity by a bi-isotropic dipolar particle, (e.g. in particular a chiral one) in an arbitrary, homogeneous or inhomogeneous, embedding medium. Also in FRET observations on transmission of energy and helicity between chiral molecules, and in its consequences for the torque exerted on each other \cite{muka,MNV2015}. In addition we shall see below that Eq.(\ref{todip3}) constitutes the basis for introducing  a new dissymmetry factor in circular dichroism and enantiomeric discrimination. Hence this new law  gives rise to  avenues worthy of further research.

\section{Consequences for the polarizabilities}
It will be  is useful to consider a Cartesian framework where  the elliptically polarized incident plane wave  has ${\bf s}_i$ along $OZ$, expressing  its electric vector in an helicity basis ${\bm \epsilon}^{\pm}=(1/\sqrt{2})(1, \pm i, 0)$ as the sum of a left-hand (LCP)  and a right-hand  (RCP) circularly polarized plane wave,  so that ${\bf e}_{i}=( e_{i x},e_{i y},0)=e_{i}^{+}{\bm \epsilon}^{+}+ e_{i}^{-}{\bm \epsilon}^{-}$ and  ${\bf  b}_{i}= ( b_{i x},b_{i y},0)=n ( -e_{i y},e_{i x},0)=b_{i}^{+}{\bm \epsilon}^{+}+ b_{i}^{-}{\bm \epsilon}^{-}=-ni(e_{i}^{+}{\bm \epsilon}^{+}- e_{i}^{-}{\bm \epsilon}^{-})$. The upper and lower sign of  $\pm$ standing for LCP (+) and RCP (-), respectively. In this representation, the incident helicity density  reads:    $ {\hel^{i}}=(\epsilon/k)\Im[e_{i x}^{*}e_{i y}]= (\epsilon/2k) S_{3}=(\epsilon/2k)[|e_{i}^{+}|^2-|e_{i}^{-}|^2]$, namely, it is the difference between the LCP and RCP intensities of the field.   $S_3 = 2 \Im[e_{i x}^{*}e_{i y}]= |e_{i}^{+}|^2-|e_{i}^{-}|^2$ is the 4th Stokes parameter \cite{marston1,born}.

Using Eq.(\ref{consti})  in the above geometry, we obtain from the helicity conservation theorem (\ref{todip1}):
\be
{\cal W}_{\hel}^{a}+\frac{8\pi c k^3}{3 \epsilon}\{ \Im[\alpha_{e}^{*}   \alpha_{me} + n^2 \alpha_{m} \alpha_{em}^{*}] |e_{i}|^2  \nonumber \\
 -2k \sqrt{\frac{\mu}{\epsilon}} \Re[\alpha_{e}^{*}   \alpha_{m} -\alpha_{me} \alpha_{em}^{*}]{\hel^{i}}\}=
    \nonumber \\
\frac{2\pi c}{n}\{n  [\alpha_{em}^{R} -  \alpha_{me}^{R}] |e_i|^2 - 2 (\alpha_{e}^{I} +n^2  \alpha_{m}^{I})  \frac{k}{\epsilon} {\hel^{i}}\}  . \label{todip2}
\ee
The superscripts $R$ and $I$ denote the real and imaginary parts of the polarizabilities, respectively.  $|e_{i}|^2= | e_{i x}|^2+|e_{i y}|^2=\frac{8\pi}{c}\sqrt{\frac{\mu}{\epsilon}} <S>=\frac{8\pi}{\epsilon} <w>$. $<S>$ and $<w>$ representing the incident field  time-averaged Poynting vector magnitude  and electromagnetic  energy density, respectively. $<w>=<w_e>+<w_m>$ .  $<w_e>=(\epsilon/16 \pi)|{\bf E}_{i}|^{2}$, $<w_m>=(1/16 \pi  \mu)|{\bf B}_{i}|^{2}$.  

In addition, in this reference frame,  the extinction torque (\ref{torque})  is: ${\bm \Gamma}= (0,0,\Gamma)=\Gamma {\bf s}_{i}$ so that the right side of Eq.(\ref{todip1}) obviously is $(4\pi c/n)\Gamma$ which is given by the right side of Eq.(\ref{todip2}), 

On the other hand, we should recall that the optical theorem for energies, Eq.(\ref{top}), leads to
\be
{\cal W}^{a}+\frac{2 k^3}{3 }\{[\epsilon^{-1} (|\alpha_{e}|^{2}    
+ n^{2}|\alpha_{em}|^2) + \mu (|\alpha_{me}|^{2}   \nonumber \,\,\,\,\,\, \\
+ n^{2}|\alpha_{m}|^2)]|e_{i}|^2  
 -4k \sqrt{\frac{\mu}{\epsilon}}\Im [ \epsilon^{-1} \alpha_{em}^{*}   \alpha_{e}  + \mu \alpha_{me} \alpha_{m}^{*}]{\hel^{i}}\}=
    \nonumber \\
2k \sqrt{\frac{\mu}{\epsilon}}(\alpha_{me}^{R} -  \alpha_{em}^{R}) {\hel^{i}}+ (\alpha_{e}^{I} +n^2  \alpha_{m}^{I}) |e_i|^2   .  \,\,\,\,\,\,\, \,\,\,\,\, \label{top2}
\ee
 Considering from now on absence of absorption from electric currents, $  {\cal W}_{\hel}^{a}=0$ and  ${\cal W}^{a}=0$. The compatibility of  the new equation (\ref{todip2})  with (\ref{top2}) implies that their combination yields
\be
\sqrt{\frac{\mu}{\epsilon}}(\alpha_{me}^{R} -  \alpha_{em}^{R})(\frac{4k^{2}}{\epsilon^{2}}{\hel^{i}}^{2}-|e_{i}|^{4})=\nonumber \\
\frac{4 k^{3}}{3 }\{(|\epsilon^{-1}\alpha_{e}-\mu \alpha_{m}|^{2}+\frac{\mu}{\epsilon}|\alpha_{em}+\alpha_{me}|^{2})\frac{k}{\epsilon}\hel^{i}|e_{i}|^{2} - \nonumber \\
\sqrt{\frac{\mu}{\epsilon}}\Im[  \alpha_{em}^{*}(\frac{4 k^{2}}{\epsilon^{2}}{\hel^{i}}^{2} \,\,\frac{\alpha_{e}}{\epsilon}-|e_{i}|^{4 } \mu \alpha_{m})+   \nonumber\\
 \alpha_{me}^{*}(|e_{i}|^{4} \,\, \frac{\alpha_{e}}{\epsilon}-\frac{4k^{2}}{\epsilon^{2}}{\hel^{i}}^{2}  \mu\alpha_{m})] \}. \,\, \, \label{cond1}
\ee
Eq.(\ref{cond1}) constitutes the constraint between the four polarizabilities $\alpha_e$, $\alpha_m$, $\alpha_{em}$ and  $\alpha_{me}$ imposed by  the conservation of the two quantities: energy and helicity.

In particular,  if the particle is not bi-isotropic, ($\alpha_{em}= \alpha_{me}=0$) and $\hel^{i}\neq 0$, the conservation of both helicity and energy, Eq.(\ref{cond1}), states that the particle is dual \cite{molina1,molina2},  i.e.  $\epsilon^{-1}\alpha_{e}=\mu \alpha_{m}$ and thus  fulfills the well-known {\it first Kerker condition} (K1) \cite{kerk,greffin} according to which it produces  zero angular distribution of scattered intensity in the backscattering direction. However, as seen next, this  also occurs for chiral particles.

Several other cases are in order, as shown next.

\subsection{Circular polarization of the incident wave}
In this case  $e_{ix}=e$ and  $e_{iy}=\pm i e$, $e$ being real, depending on whether the incident wave is LCP or RCP. Then   $|e_{i}|^2=2e^2$, and   $|e_{i}|^{4}-\frac{4k^{2}}{\epsilon^{2}}{\hel^{i}}^{2}  =0$, i.e.   $\pm |e_{i}|^{2}= \frac{2k}{\epsilon}{\hel^{i}}$ with  the sign + and - applying when  the wave is LCP and RCP, respectively. Then (\ref{cond1}) becomes
\be
\pm \{(|\epsilon^{-1}\alpha_{e}-\mu \alpha_{m}|^{2}+\frac{\mu}{\epsilon}|\alpha_{em}+\alpha_{me}|^{2}) \} =\nonumber \\
2\sqrt{\frac{\mu}{\epsilon}}\Im[ ( \alpha_{em}^{*} +  \alpha_{me}^{*}) (\frac{\alpha_{e}}{\epsilon}- \mu \alpha_{m})]. \label{condcircul}
\ee
Which yields:
\be
(\epsilon^{-1}\alpha_{e}-\mu \alpha_{m}) \pm i\frac{\mu}{\epsilon} (\alpha_{em}+\alpha_{me})=0. \label{condcircul1}
\ee
 If the particle is chiral, then \cite{tang1}   $\alpha_{em}= - \alpha_{me}$ and either (\ref{condcircul}) or  (\ref{condcircul1}) imply that  $\epsilon^{-1}\alpha_{e}=\mu \alpha_{m}$, i.e.  the particle is dual and thus holds K1.

Reciprocally, if the particle is such that $\epsilon^{-1}\alpha_{e}=\mu \alpha_{m}$, then (\ref{condcircul}) or (\ref{condcircul1}) imply that  $\alpha_{em}= - \alpha_{me}$, namely the particle is chiral and hence dual.

In addition, in this case  ${\bf p}=\pm i n {\bf m}$ and  ${\bf b}({\bf s})=\mp n i{\bf e}({\bf s})$ [cf.  Eqs.(\ref{consti}),   (\ref{dipe}) and (\ref{dipm})] i.e. the scattered field is circularly polarized (CP) with respect to the Cartesian system of orthogonal  axes defined by the unit vectors: $({\bm \epsilon}_{\perp},{\bm \epsilon}_{\parallel}, {\bf s})$, (see Fig.1). ${\bm \epsilon}_{\perp} $ and  ${\bm \epsilon}_{\parallel}$ being respectively  perpendicular and parallel to the scattering plane $OPQ$.  I.e.: ${\bf e}({\bf s})=({\bf e}({\bf s}) \cdot {\bm \epsilon}_{\perp})(1,\pm i, 0)$ and ${\bf b}({\bf s})=( n {\bf e}({\bf s}) \cdot {\bm \epsilon}_{\perp})  (\mp i, 1,0)$.  The helicity density of the scattered field being proportional to its intensity density: ${\hel^{s}}=\pm \frac{\epsilon}{2k} |{\bf e}({\bf s})|^2$; and the flow of helicity density  (spin) being proportional to that of energy density (Poynting vector).  Then, in this case the optical theorem for the helicity (\ref{todip1}) and that for the energy (\ref{top})  are equivalent.  In fact, it is known \cite{cameron1} that for circularly polarized waves there is a mapping of the helicity to the energy. Thus  both conservation laws coincide when both the  incident wave and the scattered field (like under K1)  have circular polarization.

We then conclude that  {\it the necessary and sufficient condition for a non-absorbing bi-isotropic particle to be chiral, $ \alpha_{em}=  - \alpha_{me}$, is that $\epsilon^{-1}\alpha_{e}=\mu \alpha_{m}$, i.e. it is dual. Its scattering by a circularly polarized plane wave, which must satisfy both energy and  helicity conservation, produces a circularly polarized scattered field with zero  differential scattering cross section  in the backscattering direction. Namely, the particle satisfies the first Kerker condition}.

As stated before, if $\alpha_{em}= \alpha_{me}=0$, the conservation of helicity and energy also implies duality, namely K1 and hence zero backscattering. Even though, of course, circular polarization of the scattered field will occur when the incident plane wave is circularly polarized.

\section{Significance for helicity emission/absorption from dipolar objects.  Helicity transfer  in F.R.E.T.}
The new  law (\ref{todip3}) stating the conservation of electromagnetic helicity shows us how to determine the rate of helicity dissipation from a dipolar particle or molecule in an arbitrary environment, whether homogeneous or inhomogeneous:
\be
 \frac{d{\cal W}_{\hel}}{dt}=
2\pi c \Re \{ -\frac{1}{n^2} {\bf p}  \cdot {\bf B}^{*}({\bf r}_0) + {\bf m}\cdot  {\bf E}^{*}({\bf r}_0)  \}. \label{todip5}
\ee
${\bf r}_0$ being a point of the  particle, (which is  usually convenient to consider its center if it is e.g. a sphere).  We write: ${\bf E}({\bf r}_0)= {\bf E}_{i}({\bf r}_0)+{\bf E }_{s}({\bf r}_0)$,     ${\bf B}({\bf r}_0)= {\bf B}_{i}({\bf r}_0)+{\bf B }_{s}({\bf r}_0)$. Where now the index $i$  denote the dipole field that the particle would emit in isolation; whereas $s$ stands for the field resulting from multiple scattering with  surrounding particles  or  near objects.

Eq.(\ref{todip5})  constitutes the starting point  for future studies on  the {\it helicity decay rate}  $\gamma_{\hel}$, either between the dipole and arbitrary near bodies, or between dipolar objects; in particular in the phenomenon of {\it fluorescence resonant energy transfer} (FRET) between molecules. For the latter, one does not have to be limited to the transfer of  energy, but likewise it is possible to analyse the {\it  flow of helicity between nearby particles} with  interesting effects to disclose from the additional degrees of freedom introduced by the helicity and its flow. Such technique based on Eq.(\ref{todip5}), [recall also (\ref{todip1})], that we shall call {\it resonant helicity transfer} (RHELT), or {\it fluorescence resonant helicity transfer}  (FRHELT) when fluorescence is involved, will use the concept that we herewith coin as {\it helicity transfer rate} $\gamma_{\hel}^{DA}$ between donor  $D$ and acceptor $A$, also taking their possible  bi-isotropy  (and chirality, in particular) into account, which in analogy with energy transfer  (see e.g. \cite{novotny}), we express by
\be
\frac{\gamma_{\hel}^{DA}}{\gamma_{\hel}^{0}}=\frac{{\cal W}_{\hel}^{DA}}{{\cal W}_{\hel}^{0}}.
\ee
$\gamma_{\hel}^{0}$ and ${\cal W}_{\hel}^{0}=\frac{4\pi}{3}  \frac{k^3}{\hel^{i}} \sqrt{\frac{\mu}{\epsilon} } \Im[{\bf p} \cdot {\bf m}^{*}]$,  [cf. Eq.(\ref{qs1})], representing the helicity decay rate  and  helicity yield from the donor in absence of acceptor, respectively.  And 
\be
{\cal W}_{\hel}^{DA}=2\pi c \Re \{ -\frac{1}{n^2} {\bf p}_{A}  \cdot {\bf B}_{D}^{*}({\bf r}_A) + {\bf m}_{A}\cdot  {\bf E}_{D}^{*}({\bf r}_A) 
\ee
The subindex $D$ in the fields means that they are generated by the donor, whereas the subindices $A$ stand for the excited dipole moments and position points in the acceptor.

Increasingly investigated   structures with electric and magnetic dipoles \cite{Nieto2011,luki,peng,kiv}, and their mutual interaction  \cite{MNV2010,greffin,kiv,evly},  make (\ref{todip5}) of appealing and intriguing consequences in such future studies. The details on these quantities will be the subject of a future study. 

\section{Effects on helicity enantiomeric discrimination by circular dichroism}

The weakness of the signal in enantiomeric discrimination is well known \cite{tang1,tang2,tang3,choi}. Proposals to enhance it by acting on the helicity of the illuminating wave have been  studied, \cite{tang1,tang2}. Such enhancement comes from the use of the right side of  the optical theorem of energy conservation, Eq.(\ref{top}), which for a chiral molecule or dipolar object, leads to the so-called dissymmetry factor: \cite{barron1,schellmann,tang1}
\be
g=2\frac{{\cal W}^{+}-{\cal W}^{-}}{{\cal W}^{+}+{\cal W}^{-}}. \label{gfactor1}
\ee
Where ${\cal W}^{\pm}$ is the energy  excitation  of the object, (i.e. the molecule or particle), which  equals the dissipated energy from the illuminating wave:  $\frac{\omega}{2} \Im[{\bf p} \cdot  {\bf E}_{i}^{*} +  {\bf m}    \cdot {\bf B}_{i}^{*}]  $.  Considering for example the pair of fields: $\pm {\bf \cal E}_{i}({\bf r},t)$ and  ${\bf \cal H}_{i}({\bf r},t)$:  ${\bf \cal E}_{i}({\bf r},t)=\Re [{\bf E}_{i}({\bf r}) \exp(-i\omega t)]$ and  ${\bf \cal H}_{i}({\bf r},t)=\Re [{\bf H}_{i}({\bf r}) \exp(-i \omega t)]$, whose respective helicites are: ${\hel}^{+}$ and  ${\hel}^{-}$, with ${\hel}^{+}={\hel}= -{\hel}^{-}$,   this leads to:
\be
g=- \frac{2 k n}{\pi}\frac{\alpha_{em}^{R}}{\alpha_{e}^{I} + n^{2} \alpha_{m}^{I}} \frac{{\hel}}{<w>}. \label{gfactor2}
\ee
The term $n^2\alpha_{m}^{I}$  being negligible in cases in which  $|\alpha_{m}|<<|\alpha_{e}|$ \cite{schellmann,tang1}. It should be remarked, however, that this is not always the case, see \cite{choi,Nieto2011}. Also, if the illumination is with a CP plane wave, we have seen above  that if the particle is dipolar in the wide sense and chiral then $ \alpha_e= n^2 \alpha_m$, wich would pose a difficulty to neglect $\alpha_m$ while retaining $\alpha_e$. Thus in this case one should replace the denominator of  (\ref{gfactor2}) by $2 \pi \alpha_e  <w>$ . 

It is known that the quantity $g$  may be small for the usually employed circularly polarized illumination,  for which $\pm {\bf E}_{i}$ represents the two polarization states: LCP and RCP, respectively. Then  ${\hel}^{\pm}=\pm (4\pi n/k) <w> $    and, if one can neglect the    $ \alpha_{m}^{I}$ term,  $g$ becomes the classical dissymmetry factor:  $- 8 n^2 \alpha_{em}^{R}/ \alpha_{e}^{I}$, which may be as small as $10^{-2}$ or $10^{-6}$, depending on whether there is  electronic or vibrational excitation. By contrast  the  factor  ${\hel}/<w>$  may be large enough  to overcome the above limitation of $g$ by devising differently spatially structured illumination \cite{tang1}. 

Now on account of the optical theorem of helicity established here, Eq.(\ref{todip3}), we suggest  to use the rate of helicity excitation of the object, [cf. the right side of  (\ref{todip3})]:
\be
 -{\cal W}_{\hel}'= 2\pi c \Re \{ -\frac{1}{n^2} {\bf p}  \cdot {\bf B}_{i}^{*} + {\bf m}\cdot  {\bf E}_{i}^{*}  \}=  \nonumber \\
 -32 \pi^2 c \alpha_{em}^{R}<w> + 2 \pi c\{(\alpha_{m}^{R}- \frac{ \alpha_{e}^{R}}{n^2} ) \leh +\nonumber \\
2k \sqrt{\frac{\mu}{\epsilon} } (\alpha_{m}^{I}+\frac{\alpha_{e}^{I}}{n^2}) \hel \}. \label{wgfactor}
\ee
 Where $\pm  \leh= \Re ( \pm {\bf E}_{i}\cdot {\bf B}_{i}^{*})$.

Hence, we propose a new method with  measurements based on a  novel  {\it dissymmetry factor} $g_{\hel}$ that we introduce as: 
\be
g_{\hel}=2\frac{(-{\cal W}_{\hel}'^{+}) - (-{\cal W}_{\hel}'^{-})}{(-{\cal W}_{\hel}'^{+}) + (-{\cal W}_{\hel}'^{-})}. \label{dfactor1}
\ee
Which, rather than (\ref{gfactor2}), yields  [cf.  (\ref{wgfactor})]:
\be
g_{\hel}=-\frac{(\alpha_{m}^{R}- \frac{ \alpha_{e}^{R}}{n^2} ) \leh +
2k \sqrt{\frac{\mu}{\epsilon} } (\alpha_{m}^{I}+\frac{\alpha_{e}^{I}}{n^2}) \hel }{8 \pi \alpha_{em}^{R}<w>}.  \label{ghfactor}
\ee
{\it In contrast with $g$}, Eq.(\ref{gfactor2}), {\it  that has    the usually small factor $\alpha_{em}^{R}/(\alpha_{e}^{I} + n^{2} \alpha_{m}^{I})$,  $g_{\hel}$ is a large quantity since  it contains a term with the inverse factor:  $(\alpha_{e}^{I} + n^{2} \alpha_{m}^{I})/\alpha_{em}^{R}$.}

In common situations in which  $ \leh=0$,  Eq.(\ref{ghfactor}) becomes:
\be
g_{\hel}=-\frac{k}{4\pi n \epsilon }\frac{\alpha_{e}^{I}+ n^2{\alpha_{m}^{I}}}{\alpha_{em}^{R}}
\frac{ \hel }{<w>}.  \label{ghfactornl}
\ee
(Again the term $n^2\alpha_{m}^{I}$ may be  negligible only  in cases in which  $|\alpha_{m}|<<|\alpha_{e}|$ ).

In fact for a  circularly polarized plane wave,  $ \leh=0$ and  Eq.(\ref{ghfactor}) becomes:
\be
g_{\hel}=-\frac{k}{4\pi n  }\frac{\alpha_{e}^{I}+ n^2{\alpha_{m}^{I}}}{\alpha_{em}^{R}}\frac{ \hel }{<w>} = 
-\frac{\alpha_{e}^{I}+ n^2{\alpha_{m}^{I}}}{\alpha_{em}^{R}}. \label{ghfactornl}
\ee
Thus being of the order of the inverse of the usual dissymmetry factor, i.e. of  $g^{-1}$. In addition, the helicity factor  $\frac{ \hel }{<w>}$ common to Eq.(\ref {gfactor2}) and (\ref{ghfactornl}),  enhances  not only $g$ with "superchiral light" as shown in \cite{tang1}, but then also  $g_{\hel}$.

Because of  (\ref{who}),  (\ref{todip1}) and (\ref {todip3}) one may equally determine  a helicity dissymmetry factor   on using in (\ref{dfactor1}) the quantity ${\cal W}_{\hel}^{s}=(8\pi c k^3/3 \epsilon) \Im[{\bf p} \cdot {\bf m}^{*}]$ instead of $-{\cal W}_{\hel}'$. For particles with small imaginary parts of the polarizabilities, both dissymmetry factors are equivalent, and specially when one employs  CP illumination, both   $-{\cal W}_{\hel}'$  and ${\cal W}_{\hel}^{s}$ lead to: $g_{\hel}=\frac{2 }{n} \frac{\alpha_{e}^{I}}{\alpha_{em}^{R}}$. 

To take advantage of $g_{\hel}$,  detection should be carried out  by measuring the total scattered or radiated helicity,  rather than the energy excitation, through an experiment that involves determination of Stokes parameters, including $S_3$, when LCP is used. This, according to the above helicity optical theorem,  equals the  helicity dissipated from the incident field. In the case of an incident LCP plane wave, one may also perform the measurement by  determining the projection of the extinction optical torque on the particle according to [cf. Eqs.(\ref{todip1}), (\ref{todip}) and (\ref{torque})]
\be  -{\cal W}_{\hel}'=\frac{2\pi c}{n} \{ \Re[( {\bf p}  \times {\bf e}_{i}^{*}   ) \cdot{\bf s}_{i}] + \Re[({\bf m}\times  {\bf b}_{i}^{*}) \cdot{\bf s}_{i} \}]= \nonumber \\
\frac{4 \pi c }{n}\{n \alpha_{em}^{R}|e|^2 - (  \alpha_{e}^{I} + n^2 \alpha_{m}^{I})\frac{k}{\epsilon} \hel \} ,  
\ee
either directly through an optical force experiment, or equivalently again by  the optical theorem (\ref{todip}), to measure it through a determination of the total scattered helicity.

One should  remark that determining $ g_{\hel}$ is  an approach different to ellipsometric chiroptical spectroscopy, (see e.g. \cite{choi}), in which one does not exclusively employ helicity flows ${\cal W}_{\hel}$ as proposed here, but instead one uses  the ratio 
${\cal W}_{\hel}^{s}/{\cal W}^{s} \sim {\cal Q}_{\hel}^{s}/{\cal Q}^{s}$ of the scattering helicity cross section, Eq. (\ref {qs1}),
 %\be {\cal Q}_{\hel}^{s} =\frac{4\pi}{3}  \frac{k^3}{\hel^{i}} \sqrt{\frac{\mu}{\epsilon} } \Im[{\bf p} %\cdot {\bf m}^{*}] , \ee
 to the total scattering cross section, [or scattered energy flow (\ref{ot}), normalized to the incident one: $ |<S^{i}>|=\sqrt{\frac{\epsilon}{\mu}}(c|e_i|^2/8\pi)$], for two LCP and RCP waves: 
\be  {\cal Q}^s= \sqrt{\frac{\epsilon}{\mu}}(c\|e_i|^2/8\pi) \frac{c k^4}{3 n} [\epsilon^{-1} |{\bf p}|^2 + \mu |{\bf m}|^{2}] ,  \ee

%$\frac{\omega}{2} \Im[ {\bf p}  \cdot {\bf e}_{i}^{*}   +{\bf m}\cdot  {\bf b}_{i}^{*}] $,

However the  signal obtained by this latter procedure is weaker than the one proposed here based on the factor of Eq.(\ref{dfactor1}).

\section{Conclusions}
The optical theorem that expresses the conservation of electromagnetic helicity has been put forward, from which one can define a  helicity cross section for extinction, scattering and absorption. This equation suggests intriguing consequences both  for FRET and circular dichroism. As for the former a new technique: RHELT,  (or FRHELT if emission is due to fluorescence), based on the helicity transfer rate, is proposed, whereas for the  latter we suggest a new  procedure  employing ellipsometric measurements,  from   which a dissymmetry factor based on the emitted helicity rate, here introduced and larger than the standard one, yields  greater sensitivity.

In the emerging field of silicon photonics with magnetodielectric structures, the behavior of $g$ and $g_{\hel}$ drastically changes since then the resonant $\alpha_{m}^{I}$ associated to the excitation of the  magnetic dipole  \cite{Nieto2011,luki,greffin,peng,kiv,evly}, would dominate   $\alpha_{e}^{I}$ and  $\alpha_{em}^{R}$; thus $g$ may  diminish while $g_{\hel}$  would be enhanced. Further advances in the study of  magnetodielectric chiral objects should manifest such effects.

\section{Acknowledgments}
Work  supported by the MINECO through grants FIS2012-36113-C03-03 and FIS2014-55563-REDC.

\end{document}